\documentclass[aps,prl,twocolumn,groupedaddress,showpacs,amsmath,amssymb,nofootinbib]{revtex4-2}
\usepackage[dvips]{graphicx,color}
\usepackage{tensor}

\begin{document}

\title{Is diamagnetism really acausal?}

\author{Niclas Westerberg}
\email{Niclas.Westerberg@glasgow.ac.uk}
\author{Stephen M. Barnett}
\affiliation{School of Physics and Astronomy, University of Glasgow, Glasgow G12 8QQ, UK}

\date{\today}

\begin{abstract}
Diamagnetism, in which the magnetisation in a medium opposes the direction of an applied magnetic field, is a weak but familiar effect in a wide class of materials. Being weak it is also a linear response to any applied field. The problem is that the existence of diamagnetism is in direct conflict with the requirements of causality as embodied in the familiar Kramers-Kronig relations. Nature does not care about our confusion and diamagnetism exists and physics is constrained by the requirements of causality (that effect cannot precede its cause). This puzzle has received intermittent attention from time to time, with a variety of arguments made to resolve the paradox. None of these, no matter how plausible, reveal the mechanism that resolves the existence of diamagnetism without sacrificing causality. The full resolution is presented in this letter.
\end{abstract}

\maketitle


Place a piece of dielectric in a weak electric field and it will acquire a polarisation.  This polarisation will be
in the same direction as the applied field and serves to reduce the overall strength of the field in the 
medium \cite{Jackson}.  If a time-dependent electric field is applied then the medium will both absorb some of the energy
and also impart dispersion.  These effects are linked by the Kramers-Kronig relations which are a 
manifestation, indeed the embodiment, of causality \cite{Jackson,Kramers,Kronig,Nussenzveig,King}.  
Magnetic effects are more subtle: a weak applied
magnetic field may induce a magnetisation in the direction of the applied field (paramagnetism) or 
against it (diamagnetism)\cite{vanVleck,Peierls,Nolting}.  The existence of the latter, however, is in conflict 
with the Kramers-Kronig
relations as may readily be shown.  This conflict has been noted and commented on many times before
but, we assert, not yet resolved.  In this letter we resolve the paradox.  In doing so we determine why
magnetic and electric responses are so fundamentally different and, in particular, why there is no
dia-electric effect in nature.

Consider a linear isotropic and homogeneous magneto-dielectric medium.  If we place a magnetic 
field across it then it will respond by acquiring a magnetisation:
\begin{equation}
\label{Eq1}
{\bf M}(\omega) = \chi(\omega){\bf B}(\omega) ,
\end{equation}
where ${\bf M}(\omega)$ and ${\bf B}(\omega)$ are the (complex) Fourier components of the 
magnetisation and the applied magnetic induction, and $\chi(\omega)$ is the magnetic susceptibility
\footnote{We note that the magnetisation of often written in terms of the magnetic field ${\bf H}$ rather
than the magnetic induction, albeit with a different susceptibility, $\chi_H(\omega)$.  We could work with the magnetic field 
but find it more convenient, here, to use the magnetic induction.  It is important to note that for any frequencies 
at which the imaginary part of $\chi(\omega)$ is negative, the imaginary part of $\chi_H(\omega)$ will also
be negative, with the associated problems.}.  We would expect $\chi(\omega)$ to satisfy the Kramers-Kronig
relations and thereby ensure that the process is causal.  This is not consistent with the existence of 
diamagnetism for constant magnetic fields and hence there is a paradox to be resolved.

There are a number of elements in our analysis and for this reason let us start with a plan of attack.
We begin with a brief reminder of how diamagnetism arises and, in doing so, why it is ubiquitous.
Next we demonstrate the conflict with the Kramers-Kronig relations and, thereby, complete the presentation 
of the paradox.  Before presenting our resolution of this long-standing puzzle, we pause to re-examine 
and analyse critically the solutions and plausible physical arguments that have been proposed in the past.

To understand the origins of diamagnetism it suffices to consider an electron in the ground state of a 
hydrogen atom (the simplest material system).  In the presence of a constant magnetic field we can write 
the Hamiltonian for the electron in the form 
\begin{equation}
\label{Eq2}
H = \frac{({\bf p} + e{\bf A})^2}{2m}  +  V(r) ,
\end{equation}
where $-e$ and $m$ are the charge and mass of the electron, $V(r)$ is the Coulomb potential and 
${\bf A}$ is the vector potential.  The velocity of the electron is $\dot{\bf x} = ({\bf p} + e{\bf A})/m$ 
and it follows that the kinetic angular momentum is
\begin{equation}
\label{Eq3}
{\bf L}_{\rm kin} = {\bf x}\times{\bf p} + e{\bf x}\times{\bf A} .
\end{equation}
This means that the electron has a mean kinetic angular momentum even in the ground state, for
which $\langle{\bf x}\times{\bf p}\rangle = 0$.  We can write the vector potential in the form
${\bf A} = -\frac{1}{2}({\bf x}\times{\bf B})$, where ${\bf B}$ is the magnetic induction, and  hence
find \cite{Greenshields}
\begin{equation}
\label{Eq4}
\langle({\bf L}_{\rm kin})_z\rangle = \frac{e}{2}\langle x^2 + y^2\rangle B ,
\end{equation}
where we have chosen the $z$-axis to coincide with the direction of our magnetic field.
It then follows that the induced magnetic dipole moment is
\begin{equation}
\label{Eq5}
-e\langle ({\bf x}\times\dot{\bf x})_z\rangle = -\frac{e^2}{2m}\langle x^2 + y^2\rangle B ,
\end{equation}
whish is, essentailly, Pauli's expression \cite{vanVleck, Pauli, Feynman}.
This clearly \emph{opposes} the direction of the magnetic induction.  Hence the magnetic 
response is diamagnetic in character.

In terms of our initial relation, the magnetic response to a constant magnetic field has
the form ${\bf M}(0) = \chi(0){\bf B}(0)$ and so a diamagnetic response then corresponds
to the condition $\chi(0) < 0$.  To expose the conflict with causality we need to apply the
Kramers-Kronig relations, which we use in the form
\begin{eqnarray}
\label{Eq6}
{\rm Re}\,\chi(\omega) &=& \frac{2}{\pi}\mathbb{P}\int_0^\infty d\omega' \frac{\omega'{\rm Im}\,\chi(\omega')}{\omega'^2 - \omega^2} \nonumber \\
{\rm Im}\,\chi(\omega) &=& -\frac{2}{\pi}\mathbb{P}\int_0^\infty d\omega' \frac{\omega{\rm Re}\,\chi(\omega')}{\omega'^2 - \omega^2}  ,
\end{eqnarray}
where $\mathbb{P}$ denotes the principal part.  It is clear from the first of these that
\begin{equation}
\label{Eq7}
\chi(0) = \frac{2}{\pi}\mathbb{P}\int_0^\infty d\omega' \frac{{\rm Im}\,\chi(\omega')}{\omega}
\end{equation}
and that this must be positive, as ${\rm Im}\,\chi(\omega') > 0$ for all frequencies, which 
corresponds to the condition that the material can only attenuate and not amplify
any oscillating magnetic field.

Perhaps the first, and certainly the best known, attempt to resolve the apparent conflict between
diamagnetism and causality appears in the classic text by Landau and Lifshitz \cite{Landau}.  They noted
that the proportionality in Eq. (\ref{Eq1}) should fail at sufficiently high frequencies, when the
corresponding wavelength becomes sensitive to the fact that the matter is not continuous.
For this reason they proposed introducing a high-frequency cut-off and thereby invalidating
the Kramers-Kronig relations.  This is physically plausible but, nevertheless, unsatisfactory.
The issue with diamagnetism seems to occur at zero frequency rather than at high frequencies
and, more importantly, the same argument could be made for the dielectric response to applied
electric fields and yet there is no problem with the Kramers-Kronig relations there.  A second
suggestion is that it is necessary to take into account spatial dispersion and hence to work 
with $\chi(\omega,{\bf k})$ \cite{Piaevskii,Dirdal}.  
Again this is plausible but does not explain why this should be necessary 
for magnetic responses but not for electric ones.  Moreover, it complicates the question of temporal
causality due to the inevitable presence of reflections.  A third possibility is that the phenomenon 
of diamagnetism might be nonlinear \cite{Scheel,Raabe}, but this is at odds with the observation that diamagnetism
seems to be a linear response, albeit a very weak one.  Finally, there is the possibility that 
the passivity restriction, ${\rm Im }\,\chi(\omega) > 0$, might be too restrictive \cite{Markel}.  This is indeed 
possible and is, perhaps, all that is left but then the conflict with the Kramers-Kronig 
relations needs to be resolved.  Moreover, the question of why the same phenomenon
does not occur with electric responses needs to be addressed.

To resolve the paradox we need to invoke an idea dating back to Maxwell and Faraday, that
magnetism and electricity are inextricably linked and, moreover, that time-dependent phenomena
involve both magnetic and electric effects.  This it is the \emph{product} of the permeability and the
permittivity that must have a positive imaginary part at all frequencies.  The imaginary part of
$\chi(\omega)$ can be negative for a range of frequencies, provided that ${\rm Im}\,\varepsilon(\omega)\mu(\omega)$
is positive at these (and indeed all) frequencies.  That this is indeed the case is proven below.

The phenomenon of diamagnetism is a weak one and so it suffices to treat it using lowest-order 
perturbation theory.  Consider, for simplicity, an atom within a medium in its ground state.  We have 
in mind something like an inert gas, but our derivation is not limited to such a simple case and 
should apply generally, for example, to liquid water.  The
applied field can induce virtual transitions from the ground state and these are then directly 
related to a linear response producing an electric or a magnetic dipole moment.  The natural way
to treat this is via the multipolar expansion, in which we expand in powers of the ratio of the
atomic size to the wavelength of the exciting radiation \cite{Craig}.  At first order we have the electric-dipole
interaction.  The second order includes both the magnetic-dipole and electric-quadrupole 
interaction.  To resolve our puzzle we need to go to the next order and to include, specifically,
electric-octopole transitions.  The reason for this is that our susceptibility depends on products 
of transition matrix elements and consistency with the product of two second order contributions,
each deriving from the magnetic-dipole interaction, requires us to include a term deriving from
the product of an electric-dipole (first order) and electric-octopole (third order) term.

The electric quadrupole and octopole interaction terms include, crucially, spatial derivatives
of the electric field at the position of the atom:
\begin{eqnarray}
\label{Eq8}
H_{\rm EQuad} &=& -\sum_{ij}Q_{ij}\nabla_jE_i  \nonumber \\
H_{\rm EOct} &=& -\sum_{ijk}O_{ijk}\nabla_j\nabla_kE_i ,
\end{eqnarray}
where $Q_{ij}$ and $O_{ijk}$ are the electric quadrupole and octopole moment operators,
and the summations are carried out over the three orthogonal cartesian coordinates.  When 
rotationally averaged, as is appropriate for an isotropic medium, the transition matrix elements
acquire factors that are linear (for the electric quadrupole) or quadratic (for the electric
octopole) in the wavenumber $k$.

A lowest order perturbative calculation, involving all the key interactions, diamagnetic, 
magnetic dipole, electric dipole, electric quadrupole and electric octopole terms, leads
to simple expressions for the relative permittivity and permeability in the form
\begin{eqnarray}
\label{Eq9}
\varepsilon(\omega,k) &=& 1 + \sum_e \frac{1}{\omega^2_{eg}-\omega^2 - 2i\omega\gamma_e}  \nonumber \\
& & \qquad \times \left(\Delta^{eg}_{\rm e-dip} + k^2\Delta^{eg}_{\rm quad} - k^2\Delta^{eg}_{\rm dip-oct}\right) \nonumber \\
\frac{1}{\mu(\omega)} &=&  1 + \sum_e \left(\frac{\Delta^{eg}_{\rm dia}}{\omega^2_{eg}} - 
\frac{\Delta^{eg}_{\rm m-dip}}{\omega^2_{eg} - \omega^2 - 2i\omega\gamma_e}\right) ,
\end{eqnarray}
where the summations are over all of the excited states and $2\gamma_e$ is the
linewidth of the excited state $e$.  The quantities $\Delta^{eg}_i$ are transition strengths 
associated with the different types of interaction: for example $\Delta^{eg}_{\rm e-dip}$ is 
the result the action twice of the electric dipole interaction and is proportional to 
${\bf d}^{ge}\cdot{\bf d}^{eg}$, and $\Delta^{eg}_{\rm dip-oct}$ is associated 
with a product of the electric dipole and octopole moments.  The $\Delta^{eg}_i$
are defined here so as to be positive quantities and it is a key feature of the 
dipole-octopole term that it has a minus sign.  This derives from the pair of spatial
derivatives in the electric octopole Hamiltonian, which on spatial averaging give 
$-k^2$.  This is in contrast with the quadrupole term that provides a positive 
contribution $\sim |\nabla E|^2$.

The permittivity we have derived depends, explicitly, on the wave number and, as such, indicates
a dependence on spatial dispersion, which is a complicating feature we wish to avoid.  To remove
this we take advantage of the fact that the transverse current density appearing in Maxwell's 
equations is a combination of polarisation and magnetisation terms in the form \cite{Landau,Agranovich}
\begin{eqnarray}
\label{Eq10}
{\bf j}^\perp &=& \dot{\bf P}^\perp + \mbox{\boldmath$\nabla$}\times{\bf M}  \nonumber \\
&=& \varepsilon_0(\epsilon - 1)\omega^2{\bf A} - \frac{1}{\mu_0}\left(\frac{1}{\mu} - 1\right)k^2{\bf A} ,
\end {eqnarray}
where we have expressed the polarisation and magnetisation in terms of the vector potential (in the 
Coulomb gauge), together with the permittivity and the permeability.  It is only the current that is 
physically significant and we can use this fact to reassign some of the terms in Eq (\ref{Eq10})
so as to \emph{remove} the $k$-dependence and, with this, the spatial dispersion.  This procedure 
gives our final forms of $\varepsilon(\omega)$ and $\mu(\omega)$:
\begin{eqnarray}
\label{Eq11}
\varepsilon(\omega) &=& 1 + \sum_e\frac{\Delta^{eg}_{\rm e-dip}}{\omega^2_{eg} - \omega^2 - 2i\omega\gamma_e} \nonumber \\
\frac{1}{\mu(\omega)} &=& 1 + \sum_e \frac{\Delta^{eg}_{\rm dia}}{\omega^2_{eg}} - \nonumber \\
& & \frac{\Delta^{eg}_{\rm m-dip} + \Delta^{eg}_{\rm quad}\omega^2 - \Delta^{eg}_{\rm dip-oct}\omega^2}{\omega^2_{eg} - \omega^2 - 2i\omega\gamma_e} .
\end{eqnarray}
There are three crucial results that follow, directly, form these expressions.  The first is that they are 
compatible with the existence of diamagnetism in that at (or near) zero frequency we have
\begin{equation}
\label{Eq12}
\chi(0) = 1 - \frac{1}{\mu(0)} = -\sum_e \frac{\Delta^{eg}_{\rm dia} - \Delta^{eg}_{\rm m-dip}}{\omega^2_{eg}} ,
\end{equation}
which will be positive or negative depending on which if the diamagnetic ($\Delta^{eg}_{\rm dia}$) and
paramagnetic ($\Delta^{eg}_{\rm m-dip}$) contributions dominates.  The second point is that $\chi(\omega)$
only has poles in the lower half of the complex $\omega$ plane and hence satisfies a key requirement in the 
derivation of the Kramers-Kronig relations and is, in consequence, causal.

The third point is more subtle and is that the imaginary part of $\chi(\omega)$ is, indeed, negative for a 
range of frequencies and we readily can see that these frequencies are are determined by the 
dipole-octopole induced transitions.  It may be shown, in full generality, that the imaginary part of 
$\varepsilon(\omega)\mu(\omega)$ is positive for all frequencies and hence that the electromagnetic
wave, inevitably associated with a time-dependent field, can only decay in time as required by the
dictates of causality.  To prove this is somewhat involved, so we demonstrate it here in a simple case in
which the frequency of the wave is resonant with a single, narrow electric-dipole and also electric-octopole
allowed transition at frequency $\omega_0$.  We note that ${\rm Im}\, \varepsilon(\omega)\mu(\omega)
= |\mu(\omega)|^2 {\rm Im}\, \varepsilon(\omega)/\mu^*(\omega)$ so that our condition for causality can also
be written as ${\rm Im}\,\varepsilon(\omega)/\mu^*(\omega) > 0$:
\begin{eqnarray}
\label{Eq13}
{\rm Im}\,\frac{\varepsilon(\omega)}{\mu^*(\omega)} &\approx& \frac{\Delta^0_{\rm e-dip}}{2\omega_0\gamma_0}
\left(1 + \sum_e\frac{\Delta^{eg}_{\rm dia}}{\omega^2_{eg}}\right) 
-\frac{\omega^2_0\Delta^0_{\rm dip-oct}}{2\omega_0\gamma_0}  \nonumber \\
&>&  \frac{\Delta^0_{\rm e-dip} - \Delta^0_{\rm dip-oct}\omega^2_0}{2\omega_0\gamma_0} .
\end{eqnarray}
That this is positive follows directly from the multipolar expansion in which the electric-quadrupole interaction is
smaller than the electric-dipole interaction by a term of the order of $(a_0/\lambda)^2$, where $a_0$ is the 
Bohr radius and $\lambda$ is the wavelength of the electromagnetic wave.

The essential role of the electric octopole in diamagnetism may be revealed if we observe that the permeability
tends to unity as $\omega \rightarrow \infty$.  This is both physically necessary, as media become transparent 
at very high frequencies, but also in the derivation of the Kramers-Kronig relation.  In then follows from our expression 
for $\mu(\omega)$, Eq. (\ref{Eq11}), that 
\begin{equation}
\label{Eq14}
\sum_e\left(\frac{\Delta^{eg}_{\rm dia}}{\omega^2_{eg}} + \Delta^{eg}_{\rm quad} - \Delta^{eg}_{\rm dip-oct}\right) = 0 .
\end{equation}
Hence we can write our expression for $\chi(0)$ in the form
\begin{equation}
\label{Eq15}
\chi(0) = -\sum_2\left(\Delta^{eg}_{\rm dip-oct} - \Delta^{eg}_{\rm quad} - \frac{\Delta^{eg}_{\rm m-dip}}{\omega^2_{eg}}\right) ,
\end{equation}
which will only be negative if the dipole-octopole term is present and, also, sufficiently large to overcome the
contributions from the electric quadrupole and magnetic dipole contributions.  The comparatively weak nature
of electric octopole transitions, then, provides an alternative explanation for the observed weakness of
diamagnetism compared with stronger and more familiar magnetic effects.

We have shown that the existence of diamagnetism is, of course, not in conflict with causality, as embodied 
in the Kramers-Kronig relations.  To prove this we have invoked the Maxwellian principle that time-dependent 
magnetic phenomena cannot be considered as independent from electrical phenomena and, from this, have
established that a complete analysis of diamagnetism should include electric interactions up to octopole order. 
This leads to the conclusion that the imaginary part of the susceptibility in indeed negative for some 
frequencies if the medium exhibits diamagnetic behaviour.  This does not indicate acausality, however, as the
imaginary part of $\varepsilon(\omega)\mu(\omega)$ is always positive, corresponding to absorption at all
frequencies.

Our analysis resolves one further question, which is why there is no natural dia-electric behaviour, in which 
the induced polarisation at low frequencies would oppose the applied electric field.  It is because the
electric-dipole interaction stands alone as the highest order interaction (and therefore the strongest) in
the multipolar expansion and there is no contribution, at least at frequencies of physical interest, that can
compete with it.  The magnetic-dipole interaction, in contrast, arises at the same order as the electric-quadrupole
interaction and, more importantly, the diamagnetic interaction discussed at the beginning of this paper.
Whether or not a suitably engineered metamaterial might be designed to exhibit dia-electric behaviour remains 
an interesting problem. 

Finally, there are a number of additional subtleties to be considered and we present these in a fuller account 
to be published elsewhere.

\begin{acknowledgments}
This work was supported by Royal Commission for the Exhibition of 1851, the UK Engineering and Physical 
Sciences Research Council, grant number EP/X033015/1, and the Royal Society, grant number RSRP/R/210005.
\end{acknowledgments}


\begin{thebibliography}{10}

\bibitem{Jackson}
J. D. Jackson, \emph{Classical electrodynamics} 3rd ed. (Wiley, New York, 1998).

\bibitem{Kramers}
H. A> Kramers, Atti Cong. Intern. Fisica (Transactions of the Volta Centenary Congress) Commo 
{\bf 2}, 545 (1927).

\bibitem{Kronig}
R. de L. Kronig, J. Opt. Soc. Am. {\bf 12}, 547 (1926).

\bibitem{Nussenzveig}
H. M. Nussenzveig, \emph{Causality and dispersion relations} (Academic Press, New York, 1972).

\bibitem{King}
F. W. King, \emph{Hilbert transforms}  vol. 2 (Cambridge University Press, Cambridge, 2009).

\bibitem{vanVleck}
J. H. van Vleck, \emph{The theory of electric and magnetic suceptibilities} (Oxford University
Press, Oxford, 1932).

\bibitem{Peierls}
R. E. Peierls, \emph{Quantum theory of solids} (Oxford University Press, Oxford, 1955).

\bibitem{Nolting}
W. Nolting and A. Ramakanth, \emph{Quantum theory of magnetism} (Springer Verlag, Berlin, 2009).

\bibitem{Greenshields}
C. R. Greenshields, R. L. Stamps, S. Franke-Arnold and S. M. Barnett, Phys. Rev. Lett.
{\bf 113}, 240404 (2014).

\bibitem{Pauli}
W. Pauli, Z. Phys. {\bf 2}, 201 (1920).

\bibitem{Feynman}
R. P. Feynman, R. B. Leighton, M. Sands, \emph{The Feynman Lectures on Physics}, vol. 2 (Addison-Wesley Publishing Company, Reading, 1964)

\bibitem{Landau}
L. D. Landau and E. M. Lifshitz \emph{Electrodynamics of continuous media} (Pergammon, Oxford, 1960).

\bibitem{Piaevskii}
L. Pitaevskii, International journal of quantum chemistry {\bf 112}, 2998 (2012).

\bibitem{Dirdal}.
C. A. Dirdal and J. Skaar, Eur. Phys. J. B {\bf 91}, 131 (2018).

\bibitem{Scheel}
S. Scheel and S. Y. Buhmann, Acta Phys. Slovaca {\bf 58}, 675 (2008).

\bibitem{Raabe}
C. Raabe, S. Scheel and D.-G. Welsch, Phys. Rev. A {\bf 75}, 053813 (2007).

\bibitem{Markel}
V. A. Markel, Phys. Rev. E {\bf 78}, 026608 (2008).

\bibitem{Craig}
D. P. Craig and T. Thirunamachandran, \emph{Molecular quantum electrodynamics} (Academic Press, London, 1984).

\bibitem{Agranovich}
V. M. Agranovich and V. Ginzburg, \emph{Crystal optics with spatial dispersion, and excitons} 2nd. ed. 
(Springer Verlag, Berlin, 1984).


\end{thebibliography}
\end{document}